\documentclass[runningheads]{llncs}
\usepackage[T1]{fontenc}
\usepackage{graphicx}
\usepackage{braket}
\usepackage{graphics}
\usepackage{amssymb}
\usepackage{amsmath, framed}
\usepackage{braket}
\usepackage[ruled,vlined,linesnumbered]{algorithm2e}
\SetKwComment{tcp}{\small$\triangleright$\,}{}
\usepackage{algorithm2e}
\usepackage{epsfig}
\usepackage{amsfonts}
\usepackage{amssymb}
\usepackage{enumerate}
\usepackage{booktabs}
\usepackage{multirow}
\usepackage{tabularx}
\usepackage{array}
\usepackage{enumitem}

\usepackage{circuitikz}

\usepackage{color,soul}

\usepackage{pdflscape}
\usepackage{longtable}
\usepackage{graphicx}
\usepackage{fancyhdr} 

\newcommand{\Tr}{\text{Tr}}
\usepackage{amsmath, amssymb}
\newcommand{\Mcal}{\mathcal{M}}
\newcommand{\Hcal}{\mathcal{H}}

\newcommand{\argmax}{\operatornamewithlimits{arg\,max}}

\usepackage{ifpdf}
\usepackage{qcircuit}
\usepackage{pifont}
\usepackage{graphicx}
\usepackage{braket}
\usepackage{pifont}
\usepackage{array}
\usepackage{longtable}
\usepackage{bm}
\usepackage{hyperref}
\usepackage{color}
\usepackage{makeidx}
\usepackage{cite}
\usepackage{mathtools}
\usepackage{bbm}
\usepackage{wrapfig}

\usepackage{amsmath, amssymb}    
\usepackage{tabularx}            
\usepackage{booktabs}            

\usepackage{tikz}                
\usetikzlibrary{arrows.meta}     
\usetikzlibrary{positioning}     

\usepackage{xcolor}              
\begin{document}
\title{Quantum AGI: Ontological Foundations}
%
%
\author{Elija Perrier\inst{1}\orcidID{0000-0002-6052-6798} \and
Michael Timothy Bennett\inst{2}\orcidID{0000-0001-6895-8782} }
\authorrunning{E. Perrier et al.}
%
\institute{Centre for Quantum Software \& Information, UTS, Sydney \email{elija.perrier@gmail.com} \and
Australian National University, Canberra
\email{michael.bennett@anu.edu.au}}
\maketitle              
\begin{abstract}
We examine the implications of quantum foundations for AGI, focusing on how seminal results such as Bell's theorems (non-locality), the Kochen-Specker theorem (contextuality) and no-cloning theorem problematise practical implementation of AGI in quantum settings. We introduce a novel information-theoretic taxonomy distinguishing between classical AGI and quantum AGI and show how quantum mechanics affects fundamental features of agency. We show how quantum ontology may change AGI capabilities, both via affording computational advantages and via imposing novel constraints. 
\end{abstract}
\section{Introduction}
Over the last several decades, interest in synthesising artificial intelligence and machine learning methods with emergent quantum computing proposals has grown across multiple disciplines \cite{aaronson2013quantum,preskill_quantum_2021,preskill2025beyond,proctor2025benchmarking}. Despite continuing research into quantum machine learning \cite{schuld_machine_2021,BiamonteEtAl2017,caro_generalization_2022,cerezo_challenges_2022}, quantum game theory \cite{guo_survey_2008} and hybrid quantum-classical algorithms, the impact of seminal results in quantum foundations on AGI remains under-examined. 
Current AGI paradigms \cite{hutter2004universal,RussellNorvig2020,goertzel2014} are based upon classical ontology from classical physics and computation. This includes assumptions about objective reality, local causality, definite states, and distinguishable components. Quantum mechanics, by contrast, offers a fundamentally different ontology \cite{bell_speakable_2004,kochenspecker1967,NielsenChuang2010} due to the presence of superposition, entanglement and contextuality. Most approaches for integrating quantum mechanics and theories of intelligence treat quantum information processing (QIP) \textit{instrumentally}, as a tool or extension of AGI that itself remains situated in classical. A comprehensive, formal, treatment of how quantum foundations research specifically - including Bell's theorem \cite{Bell1964}, the Kochen-Specker theorem \cite{kochenspecker1967} - impacts proposals for quantum-specific AGI is yet to be undertaken. 
We address this gap by considering whether and in what circumstances the unique ontology of quantum mechanics is consistent with AGI in ways qualitatively different from classical systems (`quantum' AGI (QAGI)). Using comparative models of classical and quantum AGI, we contribute formal theoretical results by extending three cornerstone results of quantum foundations - Bell's Theorem, the Kochen-Specker theorem and no-cloning theorem - to AGI agents, examining their practical implications for AGI architecture, learning processes, and capabilities. We also contribute a structuring of debates around quantum mechanics and AI/AGI generally by introducing a QAGI taxonomy based upon quantum information principles of registers and channels. Appendices are available in via \cite{perrier2025qagirefs}. \vspace{0.5em}


\setlength{\intextsep}{2pt}
\setlength{\columnsep}{7pt}
\begin{wrapfigure}{r}{0.35\textwidth}   
\centering
\footnotesize               

\resizebox{0.35\textwidth}{!}{
\begin{circuitikz}[every node/.style={font=\scriptsize}]
\draw[rounded corners,fill=orange!20] (3.25,7.5) rectangle node {\scriptsize CAGI} (5,6.5);
\draw[rounded corners,fill=gray!15 ] (6.25,8.75) rectangle node {\scriptsize $E_C$} (8,7.75);
\draw[rounded corners,fill=gray!15 ] (6.25,6.25) rectangle node {\scriptsize $E_Q$} (8,5.25);
\draw (4,7.5)--(4,8.5);
\draw[->] (4,8.5)--(6.25,8.5) node[pos=.5,fill=white]{\scriptsize CTC};
\draw (6.25,8)--(4.25,8)      node[pos=.5,fill=white]{\scriptsize CTC};
\draw[->] (4.25,8)--(4.25,7.5);
\draw (4,6.5)--(4,5.5);
\draw[->] (4,5.5)--(6.25,5.5) node[pos=.5,fill=white]{\scriptsize CTQ};
\draw (6.25,6)--(4.25,6)      node[pos=.5,fill=white]{\scriptsize QTC};
\draw[->] (4.25,6)--(4.25,6.5);
\draw[->] (6.75,7.75)--(6.75,6.25) node[pos=.5,fill=white]{\scriptsize CTQ};
\draw[->] (7.5,6.25)--(7.5,7.75)   node[pos=.5,fill=white]{\scriptsize QTC};
\end{circuitikz}}
\caption{\scriptsize Classical agent (CAGI) interacting via CTC, CTQ or QTC maps with classical $E_C$ or quantum $E_Q$ environments} \label{fig:CAGI}

\vspace{1.2em}  

\resizebox{0.35\textwidth}{!}{%
\begin{circuitikz}[every node/.style={font=\scriptsize}]
\draw[rounded corners,fill=blue!15 ] (3.25,7.5) rectangle node {\scriptsize QAGI} (5,6.5);
\draw[rounded corners,fill=gray!15 ] (6.25,8.75) rectangle node {\scriptsize $E_C$} (8,7.75);
\draw[rounded corners,fill=gray!15 ] (6.25,6.25) rectangle node {\scriptsize $E_Q$} (8,5.25);
\draw (4,7.5)--(4,8.5);
\draw[->] (4,8.5)--(6.25,8.5) node[pos=.55,fill=white]{\scriptsize QTC};
\draw (6.25,8)--(4.25,8)      node[pos=.55,fill=white]{\scriptsize CTQ};
\draw[->] (4.25,8)--(4.25,7.5);
\draw (4,6.5)--(4,5.5);
\draw[->] (4,5.5)--(6.25,5.5) node[pos=.55,fill=white]{\scriptsize QTQ};
\draw (6.25,6)--(4.25,6)      node[pos=.55,fill=white]{\scriptsize QTQ};
\draw[->] (4.25,6)--(4.25,6.5);
\draw[->] (6.75,7.75)--(6.75,6.25) node[pos=.5,fill=white]{\scriptsize CTQ};
\draw[->] (7.5,6.25)--(7.5,7.75)   node[pos=.5,fill=white]{\scriptsize QTC};
\end{circuitikz}}
\caption{\scriptsize Quantum agent (QAGI) interacting via QTC, CTQ or QTQ maps.} \label{fig:QAGI}
\end{wrapfigure}


\textbf{Quantum Classical Taxonomies}. To structure our analysis, we introduce a QAGI classification taxonomy reflecting the typical demarcation in quantum information sciences between physical substrates and logical superstrates. Our taxonomy comprises four domains: Classical AGI implemented on classical hardware (CS-CAGI); CS-QAGI (simulating QIP-based AGI on a classical system); QS-CAGI (quantum hardware running classical AGI); and QS-QAGI (quantum-native AGI, with both quantum hardware and algorithms). Here CAGI constitutes algorithms satisfying classical information processing criteria. QAGI are algorithms satisfying quantum information processing criteria. A summary is set out in Table 2
in the Appendix. We focus on the foundational consequences for AGI when substrates become quantum mechanical in nature i.e. QS-QAGI. We argue that the transition from classical to quantum computational substrates is an ontological shift, not just an instrumental one \cite{ozkural2012like}. To understand these differences, we must first examine how quantum phenomena such as entanglement, non-locality, contextuality, indistinguishability, and the no-cloning principle problematise implicit ontological principles that underlie AGI.
\section{Classical versus Quantum Ontology}
\label{sec:classical_agi_ontology}
\textbf{Classical ontology}. Classical AGI frameworks, whether explicitly stated or implicitly assumed, are built upon a set of ontological assumptions derived from classical physics and computation theory. Classical ontology provides the conceptual scaffolding for defining agents, environments, states, information, and interactions. This includes: (i) the existence of objective, observer (measurement) independent states of affairs (`value definiteness' form of realism in line with Mermin \cite{mermin1980quantum}); (ii) the principle of locality (that causal influences propagate no faster than light); that systems (agents) and environments are separable (separability); (iii) that systems evolve according to deterministic or classically stochastic processes (evolution) where uncertainty is epistemic; (iv) that systems possess definite identity and properties at all times (even if they cannot be directly observed or measured) (identity); (v) that objects are persistent (or continuous) through time; (vi) that the outcome of measurements is independent of other properties measured alongside it (contextuality). \\
\\
\textbf{Quantum ontology}. Quantum ontology differs from classical ontology in each of these respects in empirically verified ways that are consequential for how QS-QAGI or QS-CAGI may be implemented. The starting point for an analysis of quantum ontology are the postulates of quantum mechanics themselves (which we set out in detail in App. C):
(i) measurement of quantum states (according to which they are identified) is stochastic; superposition states lack definitive values prior to measurement; (ii) unlike classical systems, measurement probabilistically projects superpositions onto specific outcomes according to the Born rule, fundamentally disturbing quantum states which may also decohere \cite{zurek1998decoherence}; (iii) quantum systems and environments may become entangled in ways that make their states inseparable (e.g. $\rho^{AE} \neq \rho^A \otimes \rho^E$); (iv) quantum states evolve according to specific unitary evolution (Schr\"odinger's equation) via quantum trajectories distinct from classical systems (evolution); (v) identical quantum particles exhibit indistinguishability (permutation equivalence) in ways distinct from classical particles (identity); (vi) the order of measurement for non-commuting observable operators can affect quantum measurement outcomes. See Appendix Table 1 
and discussion for comparison.
\\
\\
\textbf{Foundations}. In addition to these fundamental postulates, three seminal results - the consequences of which we expand upon below - from the field of quantum foundations illustrate the profoundly different nature of quantum ontology. The first - and perhaps foremost - of these are Bell's Theorems \cite{bell_speakable_2004,bell1975theory,ClauserHorneShimonyHolt1969,HensenEtAl2015,Aspect1982} a set of local realism impossibility results. Entangled systems may exhibit correlations that experimentally violate Bell inequalities in ways inconsistent with local hidden-variable realism (see App. J.2).
The second is the Kochen-Specker theorem (contextuality) which rules out non-contextual hidden variables: an observable’s value can (unlike in classical theories) depend upon the set of compatible observables measured alongside it. The third is the no-cloning theorem \cite{wootters_single_1982} which shows that quantum states cannot be copied, having significant consequences for AGI protocols - such as self-replication, self-modification or self-reflection anchored in quantum states.  
\vspace{-2em}
\\
\section{Quantum and Classical Agent Models}
\vspace{-1em}
To illustrate the consequences of differences in classical and quantum ontology for theories of AGI, we set out two exemplary models of CAGI and QAGI below. As we show, these differences flow through to key features of AGI, such as agent/environment interaction and learning protocols (such as induction). They also have ontological implications e.g. QAGI's capacity to exhibit diachronic identity, continuity and be distinguishable. We formalise a toy QAGI model using quantum information constructs, integrating concepts of quantum registers, states, channels, and measurements (set out in more detail in App. C
and \cite{watrous_theory_2018}). We consider \textit{systems} (agents) and \textit{environments}. These are described by \textit{registers} $\texttt{X}$ (e.g. bits) comprising information drawn from a classical alphabet $\Sigma$ (App. C.1). 
Registers may be in either classical or quantum states. Changes and information transfer in those states occur via \textit{channels} (or operators). Channels can be maps of: classical-to-classical registers (CTC), classical-to-quantum (CTQ) (encoding classical information in quantum states), quantum-to-classical (QTC) (extracting classical information from quantum states e.g. measurement) or quantum-to-quantum (coherent transformations between quantum registers). Both agents and environments are framed as registers which may be classical state sets (for CAGI) or quantum state sets (for QAGI). QAGI and environments may interact in ways that are coherent (QTQ), decohering (via CTC or QTC maps) or encoding (CTQ). As we discuss in the Appendix, this structuring can also be used to model internal cognitive architecture of proposed agents (which may have both quantum and classical components).\\
\\
\textbf{Classical AGI Models}. A prominent example of a classical AGI model that relies upon classical ontological assumptions is AIXI \cite{hutter2004universal,hutter2024introduction}.  AIXI uses Bayesian inference over all computable environments to select optimal actions. Its policy $\pi$ for action $a_t$ given history $h_{<t}=a_1e_1\dots a_{t-1}e_{t-1}$ is:
\begin{align}\label{eq:AIXI_classical_main}
 \pi^{\mathrm{AX}}_{h<t}
  = \arg\max_{a_t}
    \sum_{\nu\in\Mcal_{sol}} w_\nu(h_{<t})
    \sum_{e_t}\max_{a_{t+1}}\sum_{e_{t+1}}\dots\max_{a_m}\sum_{e_m}
        G_t
        \prod_{j=t}^{m}\nu(e_j\mid h_{<j}a_j)
\end{align}
where $e_k=(o_k, r_k)$ are percepts (observation $o_k$, reward $r_k$). The universal Bayesian mixture $\xi_U$ over a class of chronological semicomputable environments $m \in \Mcal_{sol}$ is $\xi_U(e_{1:m} || a_{1:m}) := \sum_{\nu \in \Mcal_{sol}} 2^{-K(\nu)} \nu(e_{1:m} || a_{1:m})$.
Here, $K(\nu)$ is the Kolmogorov complexity of the classical Turing Machine (CTM) describing environment $\nu$ while $G_t$ is the discounted reward. AIXI is based upon classical ontological assumptions: realism (definite environments $\nu$), locality and separability (agent interacts with a distinct environment via local actions/percepts), classical stochasticity (environments $\nu$ are classical probability measures), classical identity and continuity (the AIXI agent is a single, persistent algorithm), and non-contextuality (probabilities $\nu(\cdot|\cdot)$ are objective). While AIXI is uncomputable, approximations like AIXI \cite{legg2007} provide computable variants with similar assumptions. In information-theoretic terms, AIXI can be described as follows. Let $\mathbf M_{\!c}$ hold the complete classical history
$h_{<t}=a_1e_1\dots a_{t-1}e_{t-1}$ and $\mathbf P_{\!c}$ the current
percept.  Perception is a CTC map
$\mathcal P:\mathbf E\!\to\!\mathbf P_{\!c}$. Memory updates concatenate $\mathbf P_{\!c}$ onto $\mathbf M_{\!c}$ via another CTC map
$\mathcal U$.  The decision rule~\eqref{eq:AIXI_classical_main} is realised
by a CTC channel
$\mathcal D_\xi : \mathbf M_{\!c}\!\to\!\mathbf A_{\!c}$ whose transition
matrix encodes the argmax over Bayesian mixture $\xi_U$.  Action execution is
again CTC.  All registers and channels are therefore classical and freely
copyable. Recursion in planning is implemented by cloning $\mathbf M_{\!c}$
into a scratch register $\mathbf S_{\!c}$ to explore
branching trees of depth $m$.\\
\\
\textbf{Quantum AGI Models}. QAGI can also be framed in information-theoretically. Let the QAGI agent be associated with a quantum register \texttt{A} (internal system) and its environment with a quantum register \texttt{E}. The corresponding complex (Euclidean) Hilbert spaces are $\mathcal{H}_A$ and $\mathcal{H}_E$. The agent's internal state at time $t$ is a density operator $\rho_A^{(t)} \in \mathcal{D}(\mathcal{H}_A)$. This state may encode its history, beliefs, or internal model (e.g., its representation of $\xi_Q$). The environment is in a quantum state $\rho_E^{(t)} \in \mathcal{D}(\mathcal{H}_E)$. The total system (joint) state is $\rho_{AE}^{(t)} \in \mathcal{D}(\mathcal{H}_A \otimes \mathcal{H}_E)$. If entangled, $\rho_{AE}^{(t)} \neq \rho_A^{(t)} \otimes \rho_E^{(t)}$. The agent's action $a_t$ corresponds to a choice of quantum operation (see App. C.2): 
\begin{enumerate}[label=(\alph*)]
    \item A unitary channel $U_{a_t}$ applied to a designated part of $\mathcal{H}_E$ or $\mathcal{H}_{AE}$, described by the unitary channel $\Phi_{U_{a_t}}(X) = U_{a_t} X U_{a_t}^\dagger$. This is a quantum-to-quantum channel, preserving the quantum nature of the states.
    \item A quantum instrument channel $\mathcal{I}_{a_t} = \{ \mathcal{E}_{k}^{a_t} \}_{k \in \Gamma_{obs}}$, where $\Gamma_{obs}$ is the classical alphabet of observation outcomes. Each $\mathcal{E}_{k}^{a_t}: \mathcal{L}(\mathcal{H}_{E'}) \to \mathcal{L}(\mathcal{H}_{E'})$ is a completely positive trace-non-increasing map acting on a subsystem $\mathcal{H}_{E'} \subseteq \mathcal{H}_E$, such that $\sum_k \mathcal{E}_{k}^{a_t}$ is a trace-preserving channel. A common case is a POVM measurement $\{M_k^{a_t}\}$ where $\mathcal{E}_{k}^{a_t}(X) = M_k^{a_t} X (M_k^{a_t})^\dagger$. This is decohering quantum-to-classical channel.
\end{enumerate}
What QAGI perceives is a function of what it measures. We assume measurement is a QTC channel such that the \textit{percepts} of a QAGI $e_t=(o_t,r_t)$ are classical outputs. The observation $o_t \in \Gamma_{obs}$ is the classical outcome $k$ obtained from the instrument $\mathcal{I}_{a_t}$. The reward $r_t \in \Gamma_{rew}$ (a classical alphabet, typically $\mathbb{R}$) is a classically computed function, possibly of $o_t$ and $\rho_A^{(t-1)}$. These classical values $o_t, r_t$ are stored in classical registers. The QAGI interaction loop (detailed in App. E) 
proceeds as follows. 
Given current joint state $\rho_{AE}^{(t-1)}$, the QAGI agent chooses action $a_t$. If $a_t$ involves an instrument $\mathcal{I}_{a_t}$, an outcome $o_t=k$ occurs with probability $\Pr(k | a_t, \rho_{AE}^{(t-1)}) = \Tr[\mathcal{E}_{k}^{a_t}(\Tr_A \rho_{AE}^{(t-1)})]$. The environment state (or relevant part $\mathcal{H}_{E'}$) updates to $\rho_{E'}^{(t)} = \mathcal{E}_{k}^{a_t}(\Tr_A \rho_{AE}^{(t-1)}) / \Pr(k)$. The agent updates its internal state $\rho_A^{(t-1)} \to \rho_A^{(t)}$ based on classical information $a_t, o_t, r_t$. This internal update is itself a quantum channel (possibly identity if no internal quantum state change, or more complex if internal beliefs are quantum). This model 
already highlights significant departures from classical POMDP RL agents. Moreover, maintaining quantum coherence is challenging. Decoherence timescales in current quantum hardware (microseconds to milliseconds) are much shorter than typical agent-environment interaction timescales, necessitating quantum error correction with substantial overhead \cite{fowler_surface_2012,gottesman_introduction_2010}. \\
\\
\textbf{Interaction protocols}
Using this quantum/classical channel formalism, we can distinguish how CAGI and QAGI interact differently according to different channels depending on whether the environment is classical or quantum:
\begin{enumerate}
    \item \textit{CAGI Interaction}. In Fig. \ref{fig:CAGI} the
CAGI is constituted by only
classical registers.  
Interacting with the classical environment \(E_{C}\) constitutes a CTC channel:  
the agent emits bit-valued actions, receives bit-valued percepts, and the
environment evolves under the kernel
\(P(e_{t+1}\!\mid e_{t},a_{t})\).
If the same agent is coupled to a quantum environment \(E_{Q}\) its sole
abilities are to prepare a quantum control state (CTQ) and to read back a
classical measurement record (QTC). All percepts are produced by a
decohering QTC map, and CAGI never stores qubits.
\item \textit{QAGI Interaction}. Figure \ref{fig:QAGI} shows a
QAGI whose memory is entirely
quantum.  
When QAGI interacts with a classical environment \(E_{C}\) it encodes an
action via a CTQ channel, converting the returning
bit-string into its quantum memory (a quantum register state). Processing inside the QAGI is via QTQ channels i.e. it is a form of coherent (unitary) quantum state evolution.
Interaction with the quantum environment and QAGI may occur coherently via QTQ, represented via the agent
or environment applying unitary transformation to the other (and/or becoming entangled); or via completely-positive trace-preserving (CPTP) maps \cite{watrous_theory_2018} on
\(\mathcal H_{A}\!\otimes\!\mathcal H_{E}\). Classical data arise only if the
agent or environment apply a QTC which outputs classical data while tracing out \(\rho_{AE}\).
\end{enumerate}
Thus we see two forms of learning interaction: (a) coherent learning, where the QAGI and environment (quantum) coherently evolve in a way such that when queried, the quantum register of the QAGI meets some threshold criteria for having learnt; and (b) where QAGI operates on the environment (or in later models on internal structures) in a way that allows for updating of \textit{classical} parameters via some optimisation strategy which are then re-encoded into quantum states via CTQ channels.

\subsection{Ontological Consequences}\label{sec:ontologicalconsequences}
Using the QIP-based model of CAGI and QAGI, we can now study the consequences of the shift from classical to quantum ontology that arises when computation is undertaken on quantum substrates. 
\\
\\
\textbf{States}. AIXI assumes a set of possible environments $\nu \in \Mcal_{sol}$, each a classical chronological semicomputable probability measure. The state of the environment is classical and definite, though potentially unknown to the agent. The agent's knowledge is encoded in its history $h_{<t}$ and $\xi_U$. This relies on the classical ontological assumption of \textit{realism} – that definite states and properties exist independently of observation. A QAGI's internal state is a density operator $\rho_A^{(t)} \in \mathcal{D}(\mathcal{H}_A)$, and the environment's state is $\rho_E^{(t)} \in \mathcal{D}(\mathcal{H}_E)$ which can be superpositions. Measurement outcomes are probabilistic (no definite values to all observables simultaneously if non-commuting) (Cor. \ref{thm:ks_qagi}). 
\\
\\
\textbf{Agent/Environment Interactions}. AIXI interacts with its environment via a sequence of classical actions $a_t$ and receives classical percepts $e_t = (o_t, r_t)$. The environment $\nu$ evolves according to $\nu(e_{1:m} || a_{1:m})$. This assumes \textit{locality} (actions have local effects) and \textit{separability} (the agent and environment are distinct entities with well-defined interfaces). Observations are assumed to passively reveal properties of the environment without altering its fundamental nature (beyond the causal effect of the action). A QAGI's action $a_t$ can be a quantum operation (e.g., a unitary $U_{a_t}$ or a QTC instrument $\mathcal{I}_{a_t} = \{ \mathcal{E}_{k}^{a_t} \}_k$) applied to the environment $\rho_E$ or the joint system $\rho_{AE}$. Percepts $o_t, r_t$ are classical outcomes derived from quantum measurements. Critically, measurement causes back-action, altering the environment state: $\rho_{E'}^{(t)} = \mathcal{E}_{k}^{a_t}(\Tr_A \rho_{AE}^{(t-1)}) / \Pr(k)$. Agent and environment can become entangled ($\rho_{AE}^{(t)} \neq \rho_A^{(t)} \otimes \rho_E^{(t)}$), meaning their states are no longer separable and can exhibit non-local correlations (Cor. \ref{thm:bell_qagi}). 
\\
\\
\textbf{Identity, Individuation, Continuity}. AIXI is conceived as a single, persistent algorithmic entity with enduring identity with classically distinguishable and addressable components (e.g., memory registers, processing units). This relies on classical notions of \textit{identity} and \textit{distinguishability}. If a QAGI utilises identical quantum subsystems (e.g., qubits in a register), they are subject to fundamental indistinguishability (Cor. \ref{thm:identity_qagi}). Their individual identities are subsumed by the (anti-)symmetrized joint state $P_{ij}\ket{\Psi} = \pm \ket{\Psi}$. The No-Cloning Theorem (Cor. \ref{thm:nocloning_qagi}) prevents the perfect replication of an arbitrary unknown quantum state $\rho_A$, affecting self-replication of QAGI. Identity (continuity through time) contingent on persistence of quantum states is problematised when $\rho_A$ is measured or decoheres. 
\\
\\
\textbf{Learning \& Induction}. AIXI learns by updating $\xi_U$ over classical environments based on its history $h_{<t}$. This is a form of Solomonoff induction \cite{solomonoff1964}, relying on classical probability theory and the assumption of objective, definite observations. A QAGI updates its internal state $\rho_A^{(t)}$ (or a model $\xi_Q$) based on classical percepts $a_t, o_t, r_t$ obtained from quantum measurements. The act of observing to learn changes the environment state.
Additionally, information learned about QAGI or its environment via measurement are contextual: what the QAGI is (or how information is updated) is then context and trajectory-dependent in ways unlike classical interaction. Updating a quantum belief state $\rho_A$ would involve quantum state update rules reliant on the Born rule, which differ from classical Bayesian conditioning. To fully characterise an unknown environmental quantum state $\rho_E$ (i.e., to learn it), resource-intensive quantum tomography is required, which necessitates measurements on an ensemble of identically prepared (and evolved) copies. 
\\
\\
\textbf{Sampling \& Search}. Expectations in AIXI's policy (Eq. \ref{eq:AIXI_classical_main}), $\sum_{e_t \dots e_m} (\dots)$, are over classical probability distributions derived from $\xi_U$ and may be approximated via classical sampling. Search for the optimal action $\argmax_{a_t}$ is over a classical action set. To estimate an expectation value, e.g., $\Tr(O \rho_E)$, a QAGI could potentially use quantum amplitude estimation \cite{brassard2000} which may require coherent oracle access to operations preparing $\rho_E$ and applying $O$ and repeated experiments as each sample consumes a state preparation. 
While quantum mechanics offers potential speedups (e.g. Grover's algorithm), sample and query complexity may have resource considerations compared to classical sampling from a fixed distribution.\\
\\
\textbf{Recursion \& Self-Reference}. AIXI is defined via Universal Turing Machines, which inherently support recursion. Classical information (like its own code or history $h_{<t}$) can be freely copied and inspected for recursive processing or self-referential reasoning. By contrast, a QAGI's internal program or model $\xi_Q$ is encoded in an unknown quantum state $\rho_A$ cannot be copied for recursive calls or direct self-inspection (Cor. \ref{thm:nocloning_qagi}). Thus quantum measurement problematises whether self-reference (or interaction) is possible. The requirement of unitary channels to preserve quantum states means operations are reversible, and information cannot be arbitrarily erased as in classical stack-based recursion. These differences may constrain how a QAGI could implement recursion or self-reference coherently.\\
\\
\textbf{Adaptivity \& Self-Modification}. AIXI adapts by refining $\xi_U$. More explicit self-modification, as in G\"odel machines \cite{schmidhuber2003}, involves classical proof systems and the ability to copy and replace its own code (a classical string). A QAGI adapts by updating its internal state $\rho_A^{(t)}$ or its model $\xi_Q$. It cannot simply read out its unknown quantum program, modify it, and write it back. Any modification of $\rho_A$ must occur via permissible quantum operations (channels), potentially driven by interactions, external parametrised registers (as are commonly updated in quantum machine learning) or external control. Self-modification for a QAGI operating on its own quantum information is therefore different from the classical ability to manipulate and replace explicit symbolic code.
%
%
\section{Foundational Implications for QAGI}
\label{sec:core_theorems_discontinuity}
The ontological shifts introduced by quantum mechanics lead to formal constraints on AGI and motivate the following QAGI corollaries derived from quantum foundations results. We consider these below. Detailed working is set-out in the Appendix.
\begin{corollary}[Contextual State Representation in QAGI]
\label{thm:ks_qagi}
Let a component of QAGI be described by a Hilbert space $\Hcal$ with $\dim(\Hcal) \geq 3$. The Kochen-Specker theorem \cite{kochenspecker1967} means it is impossible to assign a definite, non-contextual classical truth value (True/False, or 1/0) to all propositions $P_i$ (represented by projection operators) about the component's state such that: (i) the values are consistent with functional relations between compatible propositions (e.g., if $P_1, P_2, P_3$ are orthogonal projectors summing to $I$, exactly one must be True), and (ii) the truth value of $P_i$ is independent of the set of compatible propositions chosen for evaluation (the context).
\end{corollary}
\textit{Implications}. The internal knowledge base or state description of a QAGI component cannot, in general, be represented as a list of classical facts holding true or false independently of how the system is queried or interacts. The state of an internal variable may only become definite relative to a specific computational process (measurement context) - which is fundamentally different from CAGI.

\begin{corollary}[Non-Local State of Entangled QAGI Components]
\label{thm:bell_qagi}
If two or more components (or an agent and its environment) of a QAGI system are in an entangled quantum state $\rho_{AB} \neq \sum_k p_k \rho_A^k \otimes \rho_B^k$, then the description of their joint state may exhibit non-local correlations that violate local realism, as bounded by Bell-type inequalities \cite{bell_speakable_2004}. These correlations are instantaneous irrespective of spatial separation of the components.
\end{corollary}
\textit{Implications}. The concept of an entangled QAGI as a strictly localised entity with a separable state from its environment is problematised. Distributed QAGI components can possess shared information not attributable to local states. While this does not permit faster-than-light signalling, it implies a fundamentally correlated nature for entangled QAGI systems, affecting notions of agent boundaries and distributed information processing.

\begin{corollary}[No-Cloning Constraint on QAGI Information]
\label{thm:nocloning_qagi}
By \cite{wootters_single_1982}, there exists no universal quantum operation (unitary evolution $U$ on a composite system $\Hcal_S \otimes \Hcal_B \otimes \Hcal_A$, where $S$ is the source state, $B$ is a blank target state, $A$ is ancillary) that can perfectly copy an arbitrary unknown quantum state $\ket{\psi}_S \in \Hcal_S$ of a QAGI component to another component $\ket{0}_B \in \Hcal_B$, such that $U(\ket{\psi}_S \otimes \ket{0}_B \otimes \ket{\text{ancilla}}) = \ket{\psi}_S \otimes \ket{\psi}_B \otimes \ket{\text{ancilla}'}$ for all $\ket{\psi}_S$.
\end{corollary}
\textit{Implications}. A QAGI cannot perfectly replicate unknown parts of its internal quantum state (e.g., its belief state if encoded quantumly) or arbitrary quantum inputs. This contrasts with classical systems where information can be copied freely. This affects learning (requiring ensembles for state estimation), memory backup, and exploration.

\begin{corollary}[Indiscernibility of Identical Components]
\label{thm:identity_qagi}
Let a QAGI utilize $N$ identical quantum subsystems whose joint state
$\ket{\Psi}\in\Hcal^{\otimes N}$ lies in the fully symmetric (or
antisymmetric) subspace, i.e.\ $P_{ij}\ket{\Psi}= \pm\ket{\Psi}$ for any
swap $P_{ij}$. Then:
(i) for any local observable $O_k$ associated with subsystem $k$, its expectation value $\bra{\Psi}O_k\ket{\Psi}$ is the same for all $k$ if $O_k$ are instances of the same type of observable for each subsystem; and
(ii) no measurement acting solely within that symmetric subspace can
assign a persistent classical label or track the individual trajectory
of a specific subsystem $k$.
\end{corollary}
\textit{Implications}. Classical AGI architectures often rely on addressable, distinct components (e.g., specific memory cells or processors). While qubits have classically addressable labels, in proposals such as quantum random-access memory \cite{giovannetti2008quantum} where the address register is kept in superposition to enable parallel queries, the quantum state lacks a definite address. Information is delocalised across multiple indices until measurement collapses the superposition. Logical qubits in quantum-error-correcting codes \cite{NielsenChuang2010} may likewise be encoded non-locally across many physical qubits (distinct from classical AGI proposals), demonstrating how quantum information naturally resists attribution to any single physical component.
\\
\\
\textbf{Identity Consequences}
These foundational results have potential consequences for the identity of QAGI. First, consider a classical register $\texttt{C}$, define a \emph{copy–observation channel} on a classical state space
  $(\Sigma,\Gamma)$ as an injective map
  $\Lambda_C : \Sigma \to \Sigma \times \Gamma,\ 
  \Lambda_C(s)=(s,f(s))$ for some read-out rule $f: \Sigma \to \Gamma$.
  Define QTC measurement channel as a map
  $\Phi_{\mathcal M}(\rho)=\sum_{k\in\Gamma}M_k\rho M_k^\dagger
  \otimes\ket{k}\!\bra{k}$ for a POVM
  $\mathcal M=\{M_k\}_{k\in\Gamma}$ on a Hilbert space $\mathcal H$. Let $\texttt{S}$ denote the physical substrate realising the agent’s
internal informational state at time~$t$. For a quantum register $\texttt{Q}$, let
$\mathcal{M}=\{M_k\}_{k\in\Gamma}$ be a POVM on $\Hcal$.
The associated CPTP map is given by $\Phi_{\mathcal{M}}:
             \rho\;\longmapsto\;
             \sum_{k\in\Gamma} M_k\rho M_k^{\dagger}\otimes\ket{k}\!\bra{k}$. Using this formulation, we can illustrate the relatively simple but consequential effects of identity that occur when shifting from CAGI to QAGI (see App. J.5 
             for discussion):
\begin{enumerate}
    \item \textit{CAGI identity}. For classical substrates, writing $\sigma_C^{(t)}\!\in\!\Sigma$ for the agent’s state,
any copy–observation channel $\Lambda_C$ is injective and has a
left inverse $\Lambda_C^{-1}$ that discards the observation
component.  Hence the post-measurement state retains a
one-to-one correspondence with $\sigma_C^{(t)}$, preserving diachronic
identity.
\item \textit{QAGI identity}. Let $\rho_Q^{(t)}\!\in\!\mathcal{D}(\Hcal)$ be the quantum state and $\Phi_{\mathcal{M}}$ any non-trivial measurement channel as above.  $\Phi_{\mathcal{M}}$ is \emph{non-injective}. Distinct pre-measurement states can yield the same joint classical–quantum record. Consequently $\Phi_{\mathcal{M}}$ admits no CPTP left inverse, so the original $\rho_Q^{(t)}$ cannot be reconstructed from outcome data and residual system alone, problematising classical non-contextual concepts of identity. Thus, while a CAGI can obtain self-knowledge through repeated introspection, QAGI faces a fundamental tension: the very act of self-observation that would confirm its identity also irreversibly changes it.
\end{enumerate}
 
\vspace{-1em}

\section{Conclusion and Future Research}
\label{sec:conclusion}
\vspace{-0.5em}
We have presented an information processing-based model of CAGI and QAGI and demonstrated how certain results from quantum foundations (Bell's Theorems, the KS Theorem and No-Cloning theorem) can problematise the classical ontology underlying typical AGI proposals. The practical consequences of these quantum foundational issues for QAGI, however, depend significantly on implementation, decoherence and error correction. Further research directions include:
\begin{enumerate}
    \item \textit{Contextual Agency}. Developing AGI architectures where knowledge representation, reasoning, and decision-making are inherently contextual and account for non-locality.
    \item \textit{Quantum Identity \& Learning}. Defining coherent agent identity and learning trajectories, especially within interpretational frameworks like Everett's many-worlds or in light of fundamental indistinguishability and no-cloning. 
    \item \textit{QAGI Resource Theories}. Quantifying how quantum resources (e.g., entanglement) enable or constrain AGI capabilities.
    \item \textit{Hybrid Models}. Reformulating notions of universal intelligence (such AIXI) in a way that is compatible with quantum computability and ontological constraints.
\end{enumerate}

\newpage
\bibliographystyle{splncs04}
\bibliography{refs-final}

\end{document}